\documentstyle[12pt,epsfig]{article}
\voffset     = - 0.10 in
\begin{document}
\thispagestyle{empty}




\vspace*{1cm}

\begin{center}
{\Large\bf
The $e / h$ Method of Energy Reconstruction 
for Combined Calorimeter
}
\end{center}

\bigskip
\bigskip

\begin{center}
{\large\bf Y.A.\ Kulchitsky, M.V.\ Kuzmin}

\smallskip
{\sl Institute of Physics, National Academy of Sciences, Minsk, Belarus}
\\
\smallskip
\& \ \  
{\sl JINR, Dubna, Russia}
\bigskip

{\large\bf V.B.\ Vinogradov}

{\sl JINR, Dubna, Russia}

\smallskip
\end{center}

\vspace*{\fill}

\begin{abstract}
The new simple method of the energy reconstruction for a combined 
calorimeter, which we called the $e/h$ method, is suggested.
It uses only the known $e/h$ ratios and the electron calibration constants 
and does not require the determination of any parameters by a minimization 
technique.
The method has been tested on the basis of the 1996 test beam data
of the ATLAS barrel combined calorimeter and demonstrated the correctness 
of the reconstruction of the mean values of energies.
The obtained fractional energy resolution is 
$[(58\pm3)\%/\sqrt{E}+(2.5\pm0.3)\%]\oplus (1.7\pm0.2)/E$.
This algorithm can be used for the fast energy reconstruction in the 
first level trigger.
\end{abstract}

\newpage
\section{Introduction}
\hspace{6mm}
The key question of calorimetry generally and hadronic calorimetry 
in particular is the energy reconstruction.
This question is especially important when a hadronic calorimeter have
a complex structure being a combined calorimeter.
Such is the combined calorimeter of the ATLAS detector 
\cite{atcol94}.

In this paper we describe the new simple method of the energy reconstruction 
for a combined calorimeter, which we called the $e/h$ method, 
and demonstrate its performance on the basis of the test beam data
of the ATLAS combined prototype calorimeter.

\section{Combine Calorimeter}
\begin{figure}[tbph]
\begin{center}
\mbox{\epsfig{figure=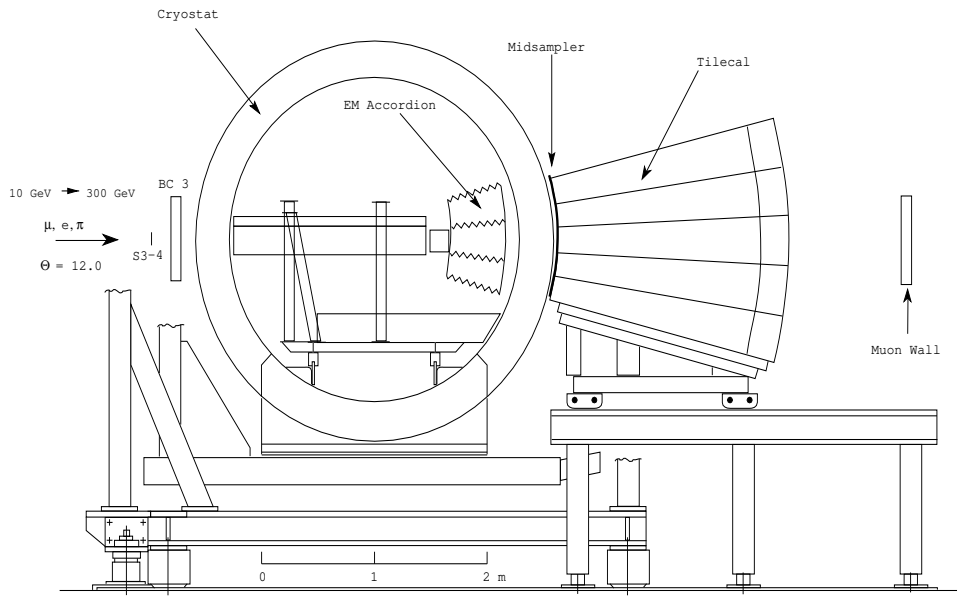,width=0.95\textwidth,height=0.4\textheight}} 
\end{center}
 \caption{
   	Test beam setup for the combined LAr and Tile calorimeters run.}
\label{fv1}
\end{figure}

\hspace{6mm}
The combined calorimeter prototype setup has been made consisting of the 
LAr electromagnetic calorimeter prototype inside the cryostat and 
downstream the Tile calorimeter prototype as shown in Fig.\ \ref{fv1}. 

The two calorimeters have been placed with their central axes at an angle 
to the beam of $12^\circ$.
At this angle the two calorimeters have an active thickness of 
10.3 $\lambda_I$.
Beam quality and geometry were monitored with a set of beam wire chambers 
BC1, BC2, BC3 and trigger hodoscopes placed upstream of the LAr cryostat.
To detect punchthrough particles and to measure the effect of longitudinal 
leakage a ``muon wall'' consisting of 10 scintillator counters (each 2 cm 
thick) was located behind the calorimeters at a distance of about 1 metre.

\subsection{Electromagnetic Calorimeter}
\hspace{6mm}
The electromagnetic LAr calorimeter  prototype
consists of a stack of three azimuthal modules, 
each one spanning $9^\circ$ in
azimuth and extending  over 2 m along the Z direction.
The calorimeter structure is defined by
2.2 mm thick steel-plated lead absorbers, folded to an accordion shape and
separated by 3.8 mm gaps, filled with liquid argon.
The signals are collected by Kapton electrodes located in the gaps.
The calorimeter extends from
an inner radius of 131.5 cm to an outer radius of 182.6 cm,
representing (at $\eta = 0$) a total
of 25 radiation lengths ($X_0$), or 1.22 interaction lengths
($\lambda_I$) for protons.
The calorimeter is longitudinally
segmented into three compartments of
$9\ X_0$, $9\ X_0$ and $7\ X_0$, respectively.
More details about this prototype can be found in 
\cite{atcol94,ccARGON}.

The cryostat has a cylindrical form
with 2 m internal diameter, filled with liquid argon, and
is made out of a  8 mm thick inner stainless-steel vessel,
isolated by 30 cm of low-density foam (Rohacell), itself protected by a
1.2 mm thick aluminum outer wall.

\subsection{Hadronic Calorimeter}
\hspace{6mm}
The  hadronic Tile calorimeter is a sampling device using
steel as the absorber and scintillating tiles as the active material 
\cite{tilecal96}.
The innovative feature of the design is the orientation of the tiles
which are placed in planes perpendicular to the Z direction 
\cite{gild91}.
For a better sampling homogeneity the 3 mm thick scintillators are 
staggered in the radial direction.
The tiles are separated along  Z  by 14 mm of steel, giving a
steel/scintillator volume ratio of 4.7.
Wavelength shifting fibres (WLS) running radially collect light from the
tiles at both of their open edges.
The hadron calorimeter prototype consists of an azimuthal
stack of five modules.
Each module covers $2\pi/64$ in azimuth and extends
1 m along the Z direction, such that the front face covers
$100\times20$\ cm$^2$. 
The radial depth, from an inner radius of 200 cm to an outer radius of 380 cm,
accounts for 8.9 $\lambda$ at $\eta = 0$\  (80.5 $X_0$).
Read-out cells are defined by grouping together a bundle of fibres
into one photomultiplier (PMT).
Each of the 100 cells is read out by two PMTs and is fully
projective in azimuth (with $\Delta \phi = 2\pi/64 \approx 0.1$),
while the segmentation along the Z
axis is made by grouping fibres into read-out cells spanning
$\Delta Z = 20$ cm ($\Delta \eta \approx 0.1$) and is therefore not projective
Each module is read out in four longitudinal segments
(corresponding to about 1.5, 2, 2.5 and 3 $\lambda_I$ at $\eta = 0$).
More details of this prototype can be found in
\cite{atcol94,ccNIM,ccrd34rep94,shower98,budagov-97-127}.

\subsection{Data Selection}
\hspace{6mm} 
Data were taken on the H8 beam of the CERN SPS, with pion and
electron beams of 20, 40, 50, 80, 100, 150 and 300 GeV.

We applied some similar to 
\cite{cobal98,comb96} 
cuts to eliminate the non-single track pion events, the beam halo,  
the events with an interaction before LAr calorimeter,
the electron  and muon events.
The set of cuts  is the following:
\begin{itemize}
\item
the single-track pion events were selected by requiring the pulse height 
of the beam scintillation counters  and the energy released in the 
presampler of the electromagnetic calorimeter to be compatible with that 
for a single particle;
\item
the beam halo events were removed with appropriate cuts on the horizontal
and vertical positions of the incoming track impact point and the space 
angle with respect to the beam axis as measured with the beam chambers;
\item
the electron events were removed by the requirement that the energy deposited 
in the LAr calorimeter is less than 90 \% of the beam energy;
\item 
a cut on the total energy rejects incoming muon;
\item
to select the events with the hadronic shower origins in the first sampling 
of the LAr calorimeter; events with the energy depositions in this sampling
compatible with that of a single minimum ionization particle were rejected;
\item  
to select the events with the well developed hadronic showers energy 
depositions were required to be more than 10 \% of the beam energy in the 
electromagnetic calorimeter and less than 70 \% in the hadronic calorimeter. 
\end{itemize}

\section{Existing Energy Reconstruction Methods}
\hspace{6mm}
Before going into description of the $e/h$ method for the energy 
reconstruction we will first briefly review the existing algorithms used 
earlier for the energy reconstruction of the ATLAS combined prototype 
calorimeter.   

Three different algorithms have been developed in order
to reconstruct the hadron energy of the ATLAS combined prototype calorimeter
\cite{cobal98,comb96}:
\begin{itemize}
\item
the benchmark method 
\cite{comb94},
\item
the sampling weighting method,
\item
the cells weighting method 
\cite{casado96}.
\end{itemize}

The benchmark algorithm  is designed to be  simple. 
With this method the incident energy is reconstructed with a minimal number of 
parameters (all energy independent with the exception of one). 
The energy of the particle is obtained as the sum of four terms: 
\begin{enumerate}
\item
The sum of the signals in the electromagnetic calorimeter, $E_{LAr}$, 
expressed in GeV using the calibration from electrons. 
\item
A term proportional to the charge deposited in the hadronic 
calorimeter, $R_{Tile}$.
\item
 A term to account for the energy lost in the cryostat, 
$E_{dm}$ $\sim$ \linebreak[4]
$\sqrt{ \vert E_{LAr, 3} \cdot a \cdot R_{Tile, 1} \vert }$. 
\item
A negative correction term, proportional to $E_{LAr}^2$. 
For showers that begin in the $EM$ calorimeter, this term crudely accounts 
for its  non-compensating behaviour.
\end{enumerate}
The  reconstructed energy  is:
\begin{equation}       
        E  = 
                E_{LAr} + a \cdot R_{Tile} 
                + b \cdot \sqrt{ \vert E_{LAr, 3} \cdot a 
		\cdot R_{Tile, 1} \vert }
                + c \cdot E_{LAr}^2 \ .
\label{eq:bench}
\end{equation}
The parameters $a$, $b$ and $c$ were determined by minimizing 
the  fractional  energy resolution $\sigma / E$ of 300 GeV pions.
However the reconstructed energy is systematically underestimated.
For this reason an additional step of rescaling is necessary.

To rescale the reconstructed energy $E$ to the beam energy $E_{beam}$ the
following expression  was used:
\begin{equation}
        E_r = 
        \frac{E}{ F_{LAr} \left[ 1 - f_h
        \left( 1- \left( \frac{\epsilon_h}{\epsilon_e} \right)_{LAr} 
        \right) \right]
        +   F_{Tile} \left[ 1 - f_h
        \left( 1- \left( \frac{\epsilon_h}{\epsilon_e} \right)_{Tile}
        \right) \right]} \ , 
\label{eq:dg2}
\end{equation}
where $f_h = 1 - f_{\pi^0}$, $f_{\pi^0}$ is the 
energy-dependent fraction of the incident hadron energy, which is
transferred to the electromagnetic sector, and given as a function 
of beam energy as in 
\cite{gabriel94}, 
two terms in the denominator are weighted by the average fractions of energy 
deposited in the LAr ($F_{LAr}$) and Tile ($F_{Tile} = 1 - F_{LAr}$)
calorimeters, taken from the 
data, the values of  $(\frac{\epsilon_h}{\epsilon_e})_{LAr}$ and
$(\frac{\epsilon_h}{\epsilon_e})_{Tile}$ (different for the two calorimeters)
were found by fitting for all beam energies and turned out to be 
$(\frac{\epsilon_e}{\epsilon_h})_{LAr}  = 1.89\pm0.04$ and
$(\frac{\epsilon_e}{\epsilon_h})_{Tile} = 1.22\pm0.02$.
As remarked in 
\cite{comb94},  
the ${\epsilon_h}$ and ${\epsilon_e}$ values in the expression (\ref{eq:dg2}) 
do not have the usual meaning as the calorimeter response to hadrons and 
electrons because they are  determined not from the raw signals but from $E$, 
which already includes corrections.
The rescaled mean values $E_r$ reasonably agreed with the nominal beam 
energies.

The sampling weighting algorithm is based on a separate correction parameter 
for each longitudinal compartment of the two calorimeters. 
These parameters are independently optimized for each incident energy and are 
indeed found to be energy-dependent.
The correction strategy  chosen here is to adjust downwards the response of 
the read-out cells with a large signal to compensate for the response to large 
EM energy clusters,  typically due to $\pi^0$ production.
A separate weighting parameter was introduced for each longitudinal
sampling. 
The energy measured in each readout cell $E_{i}$ is corrected
according to the formula: 
\begin{equation}
        E_{i}^{cor} = 
                E_{i} \cdot \Bigl( 1 - W_{j} \frac{E_i}{E_j} \Bigr) \ ,
\end{equation} 
where $E_j$ is the energy sum over all cells of sampling $j$ 
and $W_j$ is the (positive) weight to be optimized for each sampling $j$.  
In total eight energy-dependent parameters must be determined:
one for each of the seven samplings, plus an additional conversion factor
$f$ to convert the hadronic signal from charge to energy.
It turned out that the sampling weighting technique improves 
the energy resolution but does not improve the linearity 
obtained with the benchmark approach.

The cells weighting method relies on correcting 
upwards the response of cells with relatively small signals, in order 
to equalize their response to that of cells with large (typically 
electromagnetic) deposited energies.
The total energy is reconstructed correcting the energy in each   
cell of either calorimeter by a factor (typically $> 1$) which is a
function of the energy in each cell and of the beam energy.  
The reconstructed energy is expressed as:
\begin{eqnarray}
        E       = 
                  \sum_{cells} W_{em}(E_{cell},E_{beam}) \cdot  E_{cell} 
                + \sum_{cells} W_{had}(E_{cell},E_{beam}) \cdot E_{cell} 
                + E_{dm} ,
\end{eqnarray}
$W_{em}$ and $W_{had}$ are the weights to be determined.
The total number of parameters is reduced to 8 (including the cryostat 
constant).
By minimizing the energy resolution (with the constraint that the mean
reconstructed energy reproduces the nominal beam energy), a value for each
$W_i$ is found. 

In all these methods the parameters are determined by minimization.
So, it is necessary to have sufficiently large and unbiased
sample of events and to perform  the sophisticated off-line data treatment. 
Besides, for the two first methods it is necessary to rescale 
the reconstructed energy in order to obtain the nominal beam energy
values.

\section{$e / h$ Method of Energy Reconstruction}

\subsection{Algorithm}
\hspace{6mm}
The response, $R_h$ of a calorimeter to a hadronic shower is the sum 
of the contributions from the electromagnetic, $E_e$, and hadronic, 
$E_h$, parts of the incident energy 
\cite{wigmans88,groom89}
\begin{equation}
        E_{inc} = E_e + E_h \ .
\label{ev18}
\end{equation}
\begin{equation}
        R_h = 
                  e \cdot E_e + h \cdot E_h 
            =     e \cdot E_{inc} \cdot 
                 (f_{\pi^0} + (h / e) \cdot ( 1 - f_{\pi^0})) \ , 
\label{ev9}
\end{equation}
where $e$ ($h$) is the energy independent coefficient of
transformation of the electromagnetic 
(pure hadronic, low-energy hadronic activity) energy to response,
$f_{\pi^0} = E_e / E_{inc}$ is the fraction of electromagnetic energy.
From this
\begin{equation}
        E_{inc} = 
                \frac{e}{\pi} \cdot \frac{1}{e} \cdot R_h \ ,
\label{ev16}
\end{equation}
where
\begin{equation}
        \frac{e}{\pi} = 
                \frac{e/h}{1+(e/h-1)f_{\pi^0}} \ .
\label{ev10}
\end{equation}

For a combined calorimeter the beam energy, $E_{beam}$, deposits into 
the LAr compartment, $E_{LAr}$, the Tilecal compartment, $E_{Tile}$, and into 
the passive material between the LAr and Tile calorimeters, $E_{dm}$,
\begin{equation}
        E_{beam} =  E_{LAr} + E_{Tile} + E_{dm} \ . 
\label{ev13}
\end{equation}

Using the expressions (\ref{ev16}) - (\ref{ev13}) 
the following equation for the energy reconstruction has been derived:
\begin{equation}
        E =   	  c_{LAr}  \cdot (e/\pi)_{LAr}  \cdot R_{LAr} 
                + c_{Tile} \cdot (e/\pi)_{Tile} \cdot R_{Tile} 
                + E_{dm} \ ,
\label{ev7}
\end{equation}
where
\begin{equation}
	\Bigl( \frac{e}{\pi} \Bigr)_{LAr} = 
		\frac{(e/h)_{LAr}}{1 + ((e/h)_{LAr} - 1) f_{\pi^0, LAr}} \ ,
\label{epl}
\end{equation} 
 \begin{equation}
	f_{\pi^0, LAr} = k_{LAr} \cdot \ln{E} \ , 
\label{efpl}
\end{equation}
\begin{equation}
	\Bigl( \frac{e}{\pi} \Bigr)_{Tile} = 
		\frac{(e/h)_{Tile}}{1 + ((e/h)_{Tile} - 1) f_{\pi^0, Tile}} \ ,
\label{ept}
\end{equation}  
\begin{equation}
	f_{\pi^0, Tile} = k_{Tile} \cdot \ln{(E_{Tile})} \ ,
\label{fpt}
\end{equation}
\begin{equation}
\label{ev19}
        E_{dm} = c_{dm} \cdot \sqrt{E_{LAr, 3} \cdot E_{Tile, 1}} \ , 
\end{equation}
\begin{equation}
\label{el3}
        E_{LAr, 3} = c_{LAr}  \cdot (e/\pi)_{LAr}  \cdot R_{LAr, 3} 
\end{equation}
is the energy released in the third  depth of the electromagnetic calorimeter,
\begin{equation}
\label{et1}
        E_{Tile, 1} = c_{Tile} \cdot (e/\pi)_{Tile} \cdot R_{Tile, 1}
\end{equation}
is the energy released in the first depth of the hadronic calorimeter.
In order to use the equation (\ref{ev7}) it is necessary to know the values
of the following constants which have been taken from 
\cite{kulchitsky-99-303} 
and equal to:
$c_{LAr}  = 1/e_{LAr}  = 1.1$,  $c_{Tile} = 1/e_{Tile} = 0.145\pm0.002$,
$(e/h)_{LAr}  = 1.77\pm0.02$, 	$(e/h)_{Tile} = 1.3\pm0.03$,
$k_{LAr} = k_{Tile} = 0.11$,  	$c_{dm} = 0.31$.

Note that the above formulae have been represented earlier in our work 
\cite{kulchitsky-99-303}
devoted to the determination of the $(e/h)$ ratio of the LAr electromagnetic 
compartment of the ATLAS barrel combined prototype calorimeter.

\subsection{Iteration Procedure}
\hspace{6mm}
For the energy reconstruction by the formula (\ref{ev7}) it is necessary to 
know the $(e/\pi)_{Tile}$ ratio and the reconstructed energy itself.
Therefore, the iteration procedure has been developed.

Two iteration cycles were made.
The first one is devoted to the determination of the $(e/\pi)_{Tile}$ ratio.
The expression (\ref{ept}) can be written as
\begin{equation}
	\Bigl(\frac{e}{\pi}\Bigr)_{Tile} = 
		\frac{(e/h)_{Tile}}{1+((e/h)_{Tile}-1) 
		\cdot k_{Tile} 
		\cdot \ln{(c_{Tile} \cdot (e/\pi)_{Tile} \cdot R_{Tile})}} \ ,
\label{epti}
\end{equation} 
As the first approximation, the value of $(e/\pi)_{Tile}$ is
calculated using the following equation
\begin{equation}
	\Bigl(\frac{e}{\pi}\Bigr)_{Tile}^{0} = 
		\frac{(e/h)_{Tile}}{1+((e/h)_{Tile}-1) 
		\cdot k_{Tile} 
		\cdot \ln{(c_{Tile} \cdot 1.13 \cdot R_{Tile})}} \ ,
\label{epti-0}
\end{equation} 
where in the right side of this equation we used 
$(e/\pi)_{Tile} = 1.13$ corresponding to $f_{\pi^0, Tile} = 0.5$.
The obtained value $(e/\pi)_{Tile}$ in $\nu$-iteration ($\nu = 0,\ 1, \ldots$) 
are used in $(\nu +1)$-iteration in the right side of the equation 
(\ref{epti}) and the iteration continues.
The iteration process is stopped when the convergence criterion
\begin{equation}
	\mid (e/\pi)_{Tile}^{\nu + 1}-(e / \pi)_{Tile}^{\nu }\mid 
	/(e/\pi)_{Tile}^{\nu}<\epsilon    
\label{iterept}
\end{equation} 
is satisfied. 

The second iteration cycle is the determination of energy.
As the first approximation, the value of $E$ is calculated using the following 
equation with the $(e/\pi)_{Tile}$ ratio obtained in the first iteration cycle
\begin{equation}
        E_0 =     c_{LAr}  \cdot (e/\pi)_{LAr}^0  \cdot R_{LAr} 
                + c_{Tile} \cdot (e/\pi)_{Tile} \cdot R_{Tile} 
                + E_{dm} \ ,
\label{ev7-0}
\end{equation}
where
\begin{equation}
	\Bigl(\frac{e}{\pi}\Bigr)_{LAr}^0 = 
	\frac{(\frac{e}{h})_{LAr}}{1+((\frac{e}{h})_{LAr}-1) k_{LAr} 
\ln{(c_{LAr} 1.27 R_{LAr}+c_{Tile}(\frac{e}{\pi})_{Tile} R_{Tile}+E_{dm})}}
\label{epti-00}
\end{equation}   
In the right side  of this equation we used $(e/\pi)_{LAr} = 1.27$ 
corresponding to $f_{\pi^0, LAr} = 0.5 = 0.11 \ln{(100\ GeV)}$.
The convergence criterion is
\begin{equation}
	\mid E_{\nu +1} - E_{\nu} \mid / E_{\nu} < \epsilon \ .    
\label{itere}
\end{equation}
\begin{table}[tbh]
\begin{center}
\caption{
 	The average number of iterations $<N_{it}>$ 
        for the various beam energies needed to 
	receive the given value of accuracy $\epsilon$.
	} 
\label{tv3}
\begin{tabular}{|c|c|c|c|c|}  
\multicolumn{3}{c}{\mbox{~~}}\\[-2 mm]
\hline
$E$  &$<N_{it}>$      &$<N_{it}>$	&$<N_{it}>$  	&$<N_{it}>$ \\ 
(GeV)&$\epsilon=0.1\%$&$\epsilon=0.3\%$ &$\epsilon=0.5\%$&$\epsilon=1.\%$\\ 
\hline
10   & 1.79   		& 1.62 		& 1.49  	& 1.18 \\
20   & 1.63   		& 1.39 		& 1.15  	& 1.10 \\ 
40   & 1.41   		& 0.91 		& 0.78  	& 0.72 \\ 
50   & 1.28   		& 0.80 		& 0.74  	& 0.69 \\   
80   & 0.77   		& 0.68 		& 0.58  	& 0.29 \\   
100  & 0.56   		& 0.30 		& 0.18  	& 0.09 \\    
150  & 0.82   		& 0.70 		& 0.62  	& 0.39 \\    
300  & 1.01   		& 0.74 		& 0.69  	& 0.63 \\   
\hline
\end{tabular}
\end{center}
\end{table}

Table \ref{tv3} gives the average number of iterations $<N_{it}>$ for the 
various beam energies needed to receive the given value of accuracy $\epsilon$.
As can be seen, it is sufficiently only the first approximation  
for achievement 
, on average,
of convergence with an accuracy of $\epsilon = 1 \%$ 
for energies 80 -- 150 $GeV$ and 
it is necessary to perform one iteration
for the energies at 10 -- 50, 300 $GeV$.

\begin{figure*}[tbph]
\begin{center}   
\begin{tabular}{c}
\epsfig{figure=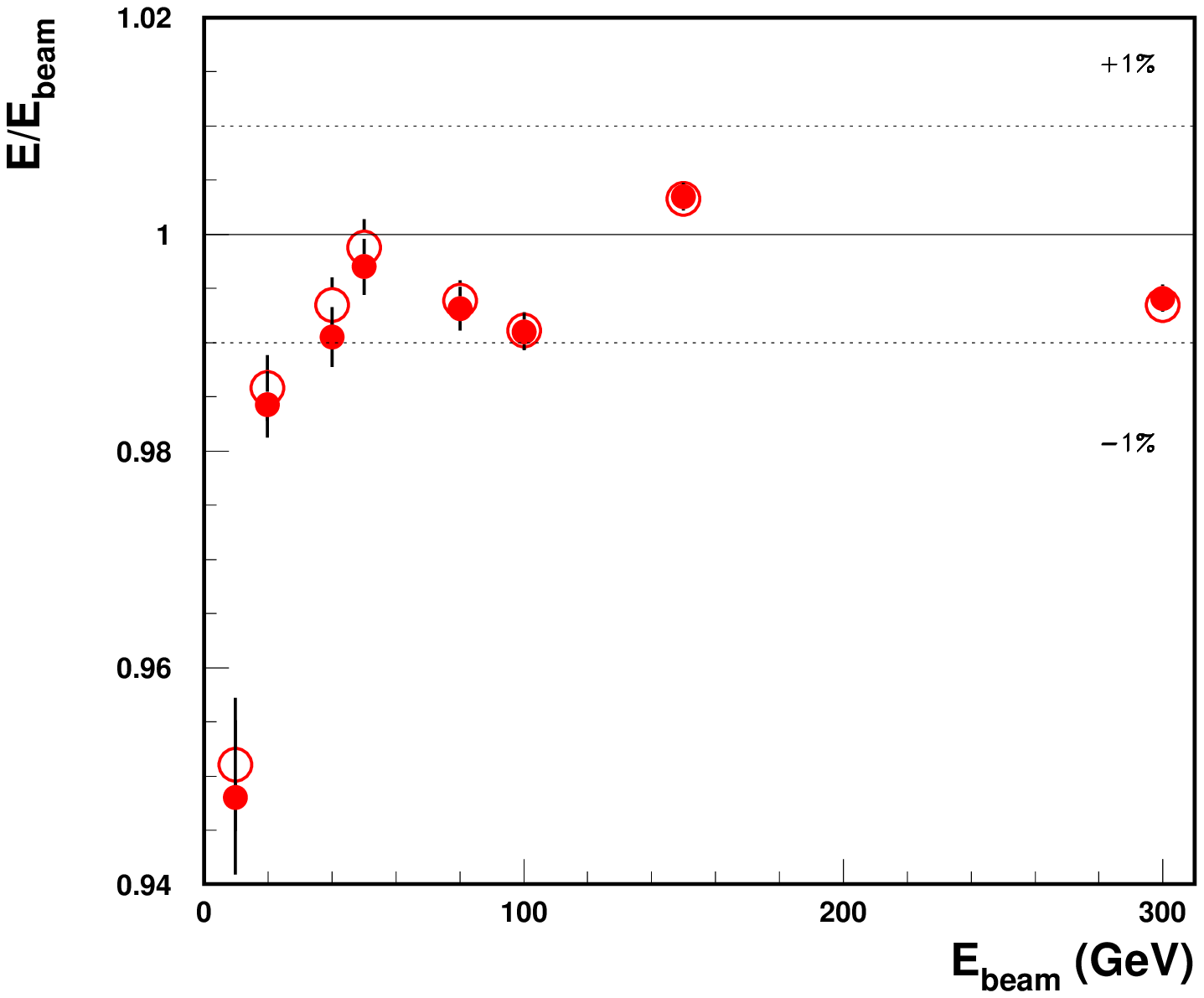,width=0.95\textwidth,height=0.9\textheight} 
\\
\end{tabular}
\end{center}
\vspace*{-20mm}
       \caption{
         Energy linearity as a function of the beam energy for 
         the $e/h$ method  obtained using the iteration procedure 
         with $\epsilon = 0.1\%$ (black circles) and the first 
	 approximation (open circles).
       \label{f03-0}}
\end{figure*}

\begin{figure*}[tbph]
\begin{center}   
\begin{tabular}{c}
\epsfig{figure=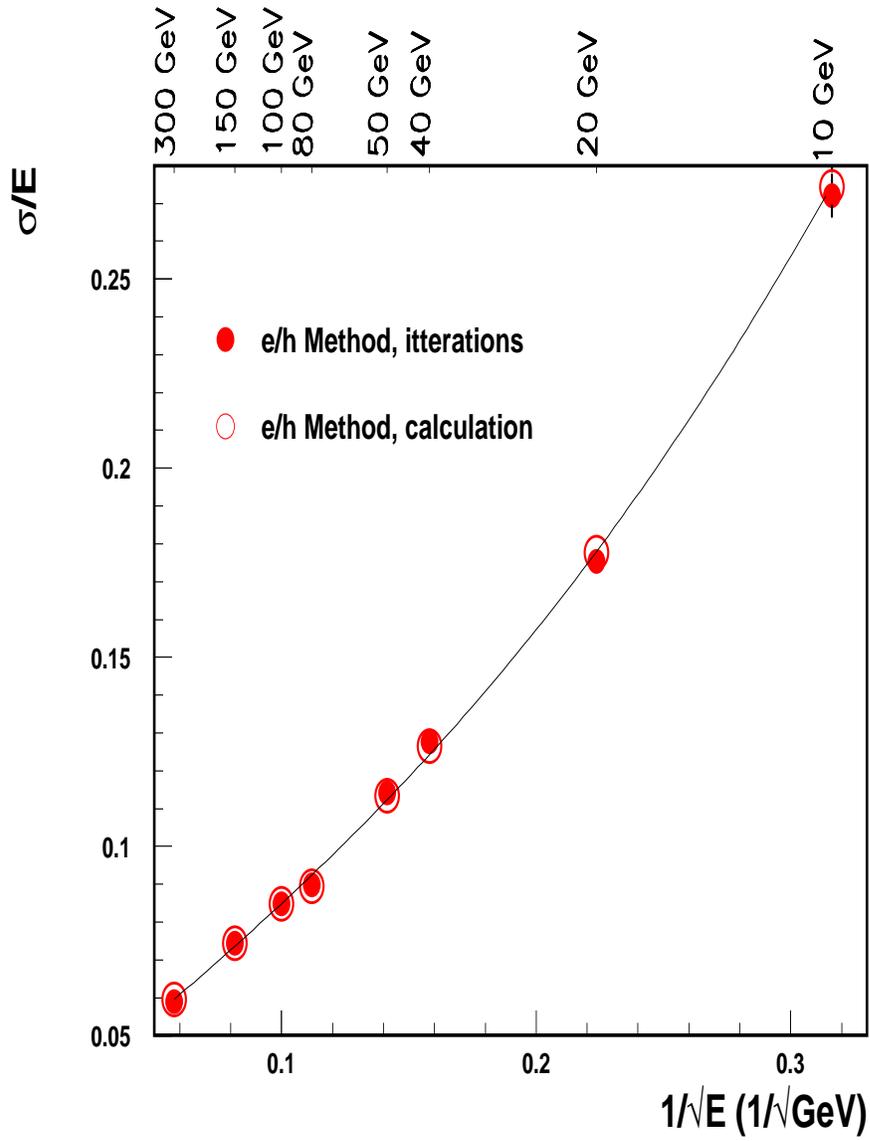,width=0.95\textwidth,height=0.9\textheight} 
\\
\end{tabular}
\end{center}
\vspace*{-20mm}
       \caption{
         The fractional energy resolutions obtained 
         with the $e/h$ method (black circles), 
         and the first approximation (open circles).          
       \label{f05-0}}
\end{figure*}

We specially investigated the accuracy of the first approximation of energy, 
$E_0$.
Fig.\ \ref{f03-0} shows the comparison between the energy linearities, the 
mean values of $E / E_{beam}$, obtained using the iteration procedure with 
$\epsilon = 0.1\%$ 
(black circles) and the first approximation 
of energy 
(open circles).
Fig.\ \ref{f05-0} shows the comparison between the energy resolutions 
obtained using these two approaches.
As can be seen, the compared values are consistent within errors.

The suggested algorithm of the energy reconstruction can be used for
the fast energy reconstruction in the first level trigger.

\begin{table}[tbh]
\begin{center}
\caption{
        Mean reconstructed energy, energy resolution and 
        fractional energy resolution for the various beam energies
        obtained using the iteration procedure with $\epsilon = 0.1\%$
	}
\label{tv1}
\begin{tabular}{|r|c|c|c|}
\multicolumn{4}{c}{\mbox{~~}}\\[-2 mm]
\hline
$E_{beam}$&$E$ (GeV)&$\sigma$ (GeV)&$\sigma / E\ (\%)$\\ 
\hline
$10^{\ast}$     GeV&  $9.30\pm0.07$& $2.53\pm0.05$&$27.20\pm0.58$\\
$20^{\star}$    GeV& $19.44\pm0.06$& $3.41\pm0.06$&$17.54\pm0.31$\\
40              GeV& $39.62\pm0.11$& $5.06\pm0.08$&$12.77\pm0.21$\\
50              GeV& $49.85\pm0.13$& $5.69\pm0.13$&$11.41\pm0.26$\\
80              GeV& $79.45\pm0.16$& $7.14\pm0.14$& $8.99\pm0.18$\\
100             GeV& $99.10\pm0.17$& $8.40\pm0.16$& $8.48\pm0.16$\\
150             GeV&$150.52\pm0.19$&$11.20\pm0.18$& $7.44\pm0.12$\\
300             GeV&$298.23\pm0.37$&$17.59\pm0.33$& $5.90\pm0.11$\\
\hline
\multicolumn{4}{@{}l@{}}{\mbox{~~}}\\[-5 mm]
\multicolumn{4}{@{}l@{}}{\underline{\mbox{~~~~~~~~~~~~~~~~~~~~~~~~}}}	\\
\multicolumn{4}{@{}l@{}}{$^{\ast}$The measured value of the 
					beam energy is 9.81 $GeV$.}	\\
\multicolumn{4}{@{}l@{}}{$^{\star}$The measured value of the 
					beam energy is 19.8 $GeV$.}	\\
\end{tabular}
\end{center}
\end{table}

\subsection{Energy Spectra}
\begin{figure*}[tbph]
\begin{center}   
\begin{tabular}{cc}
\epsfig{figure=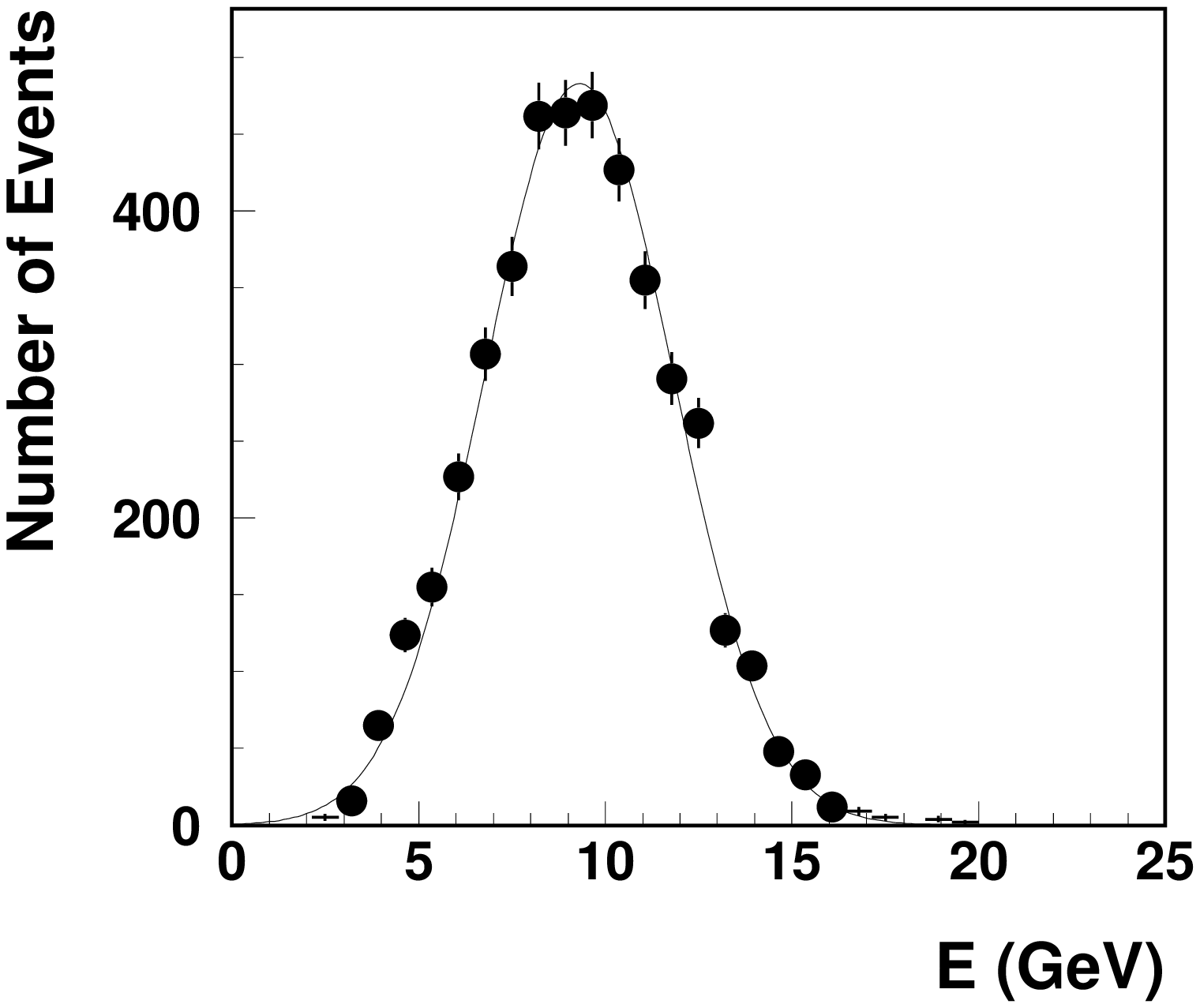,width=0.45\textwidth,height=0.4\textheight} 
&
\epsfig{figure=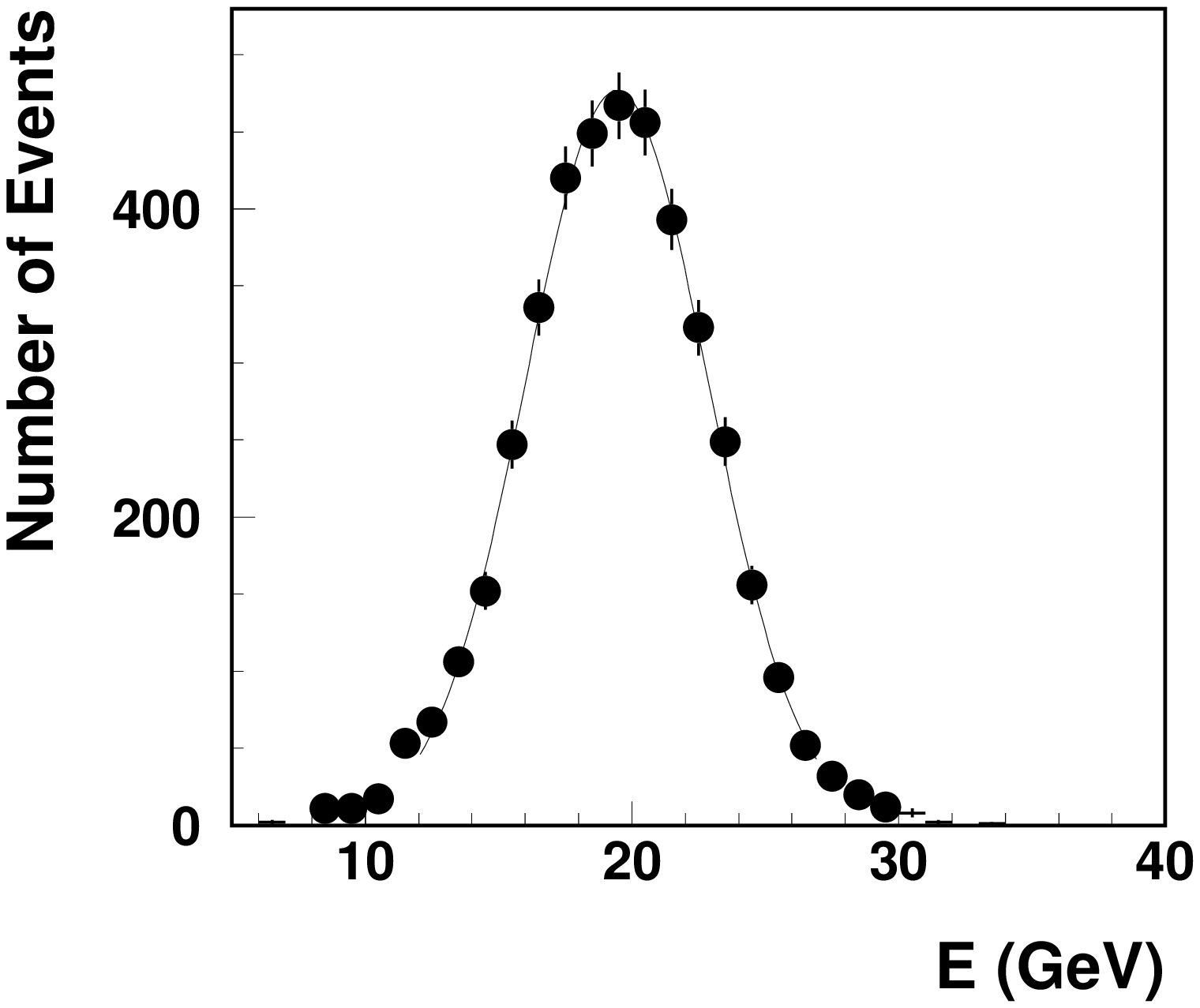,width=0.45\textwidth,height=0.4\textheight} 
\\
\epsfig{figure=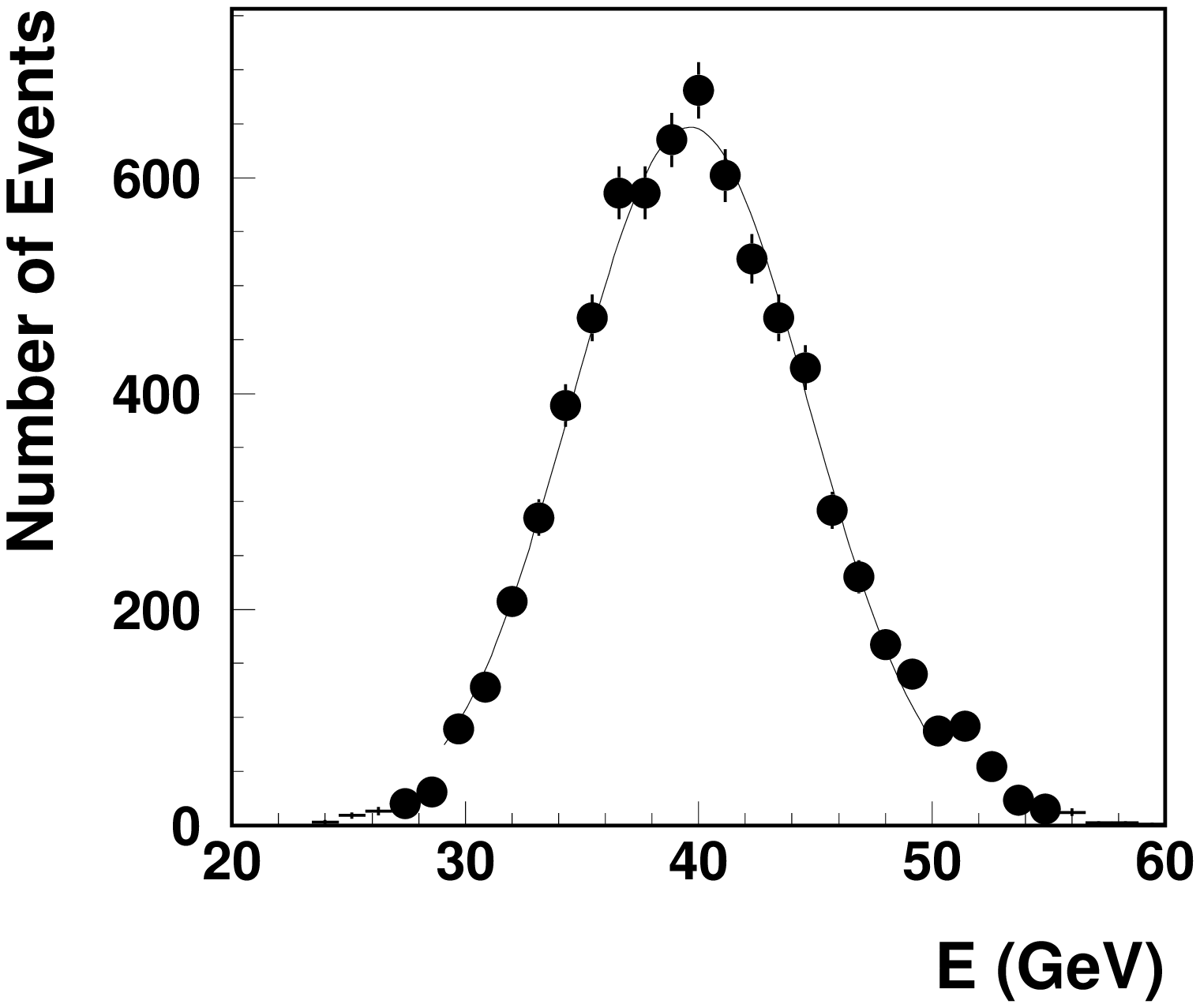,width=0.45\textwidth,height=0.4\textheight} 
&
\epsfig{figure=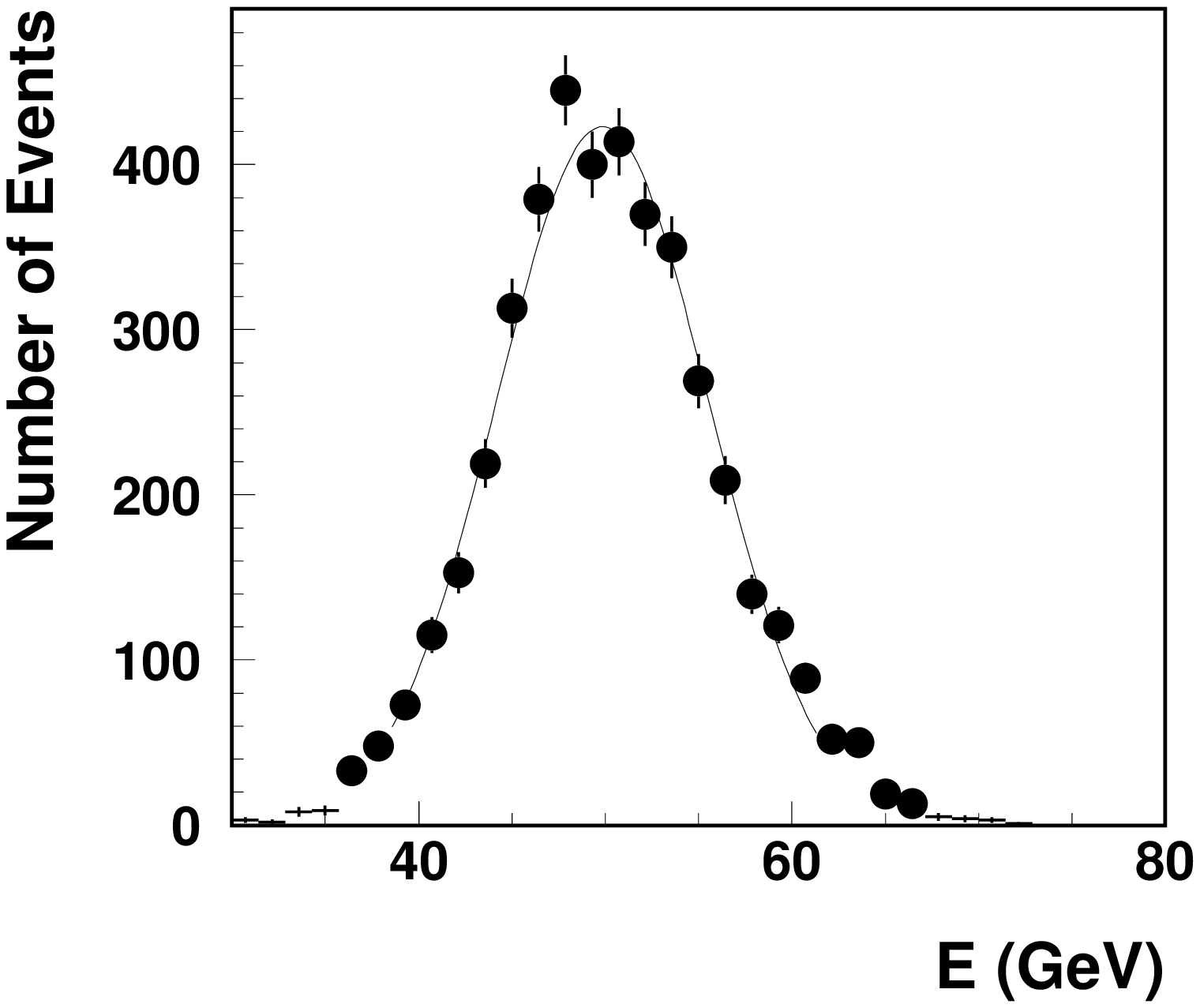,width=0.45\textwidth,height=0.4\textheight} 
\\
\end{tabular}
\end{center}
       \caption{
                The energy distributions  for $E_{beam}$ = 10, 40  GeV
                (left column, up to down) 
                and $E_{beam}$ = 20, 50  GeV (right column, up to down).  
       \label{f01}}
\end{figure*}

\begin{figure*}[tbph]
\begin{center}   
\begin{tabular}{cc}
\epsfig{figure=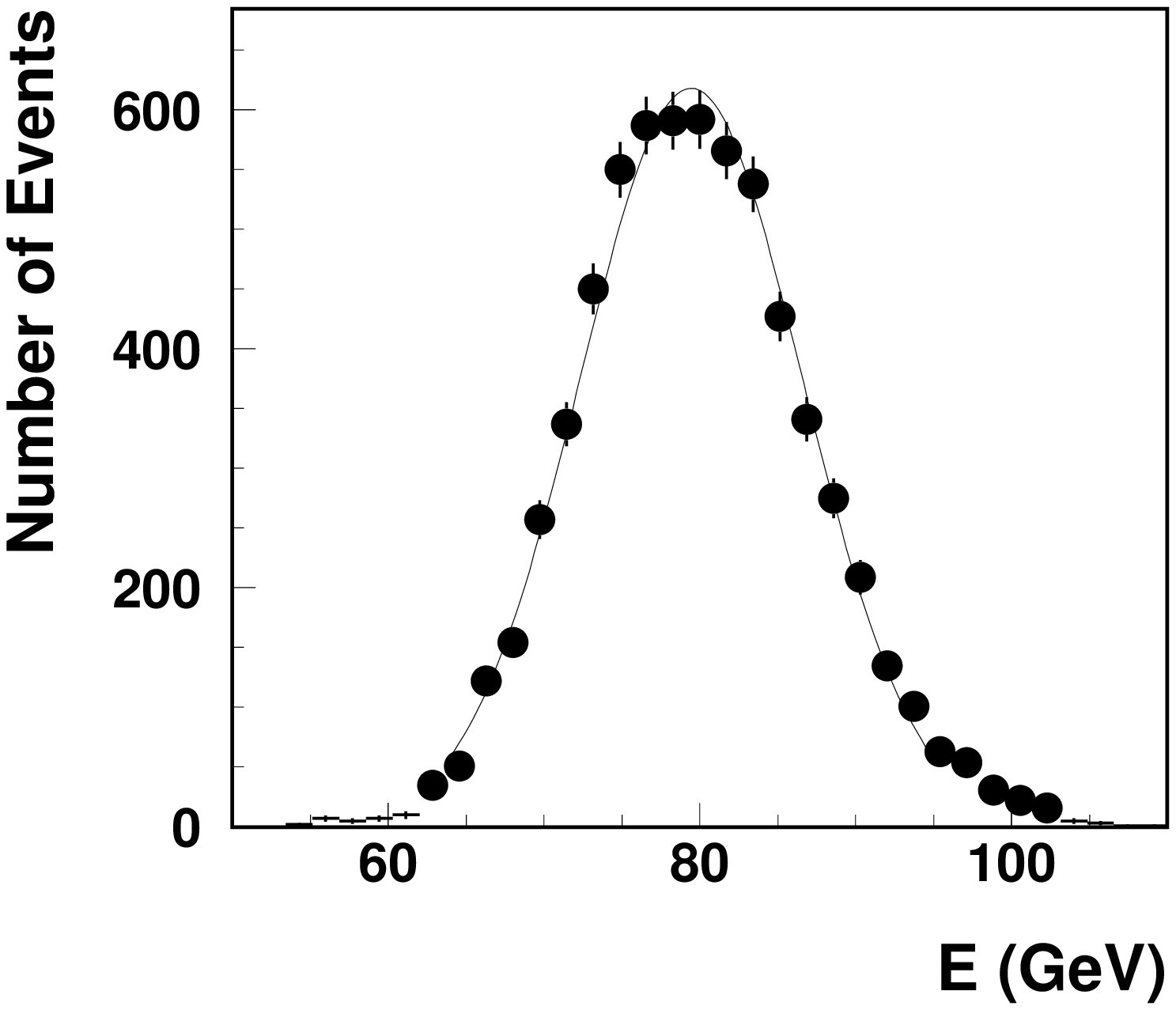,width=0.45\textwidth,height=0.4\textheight} 
&
\epsfig{figure=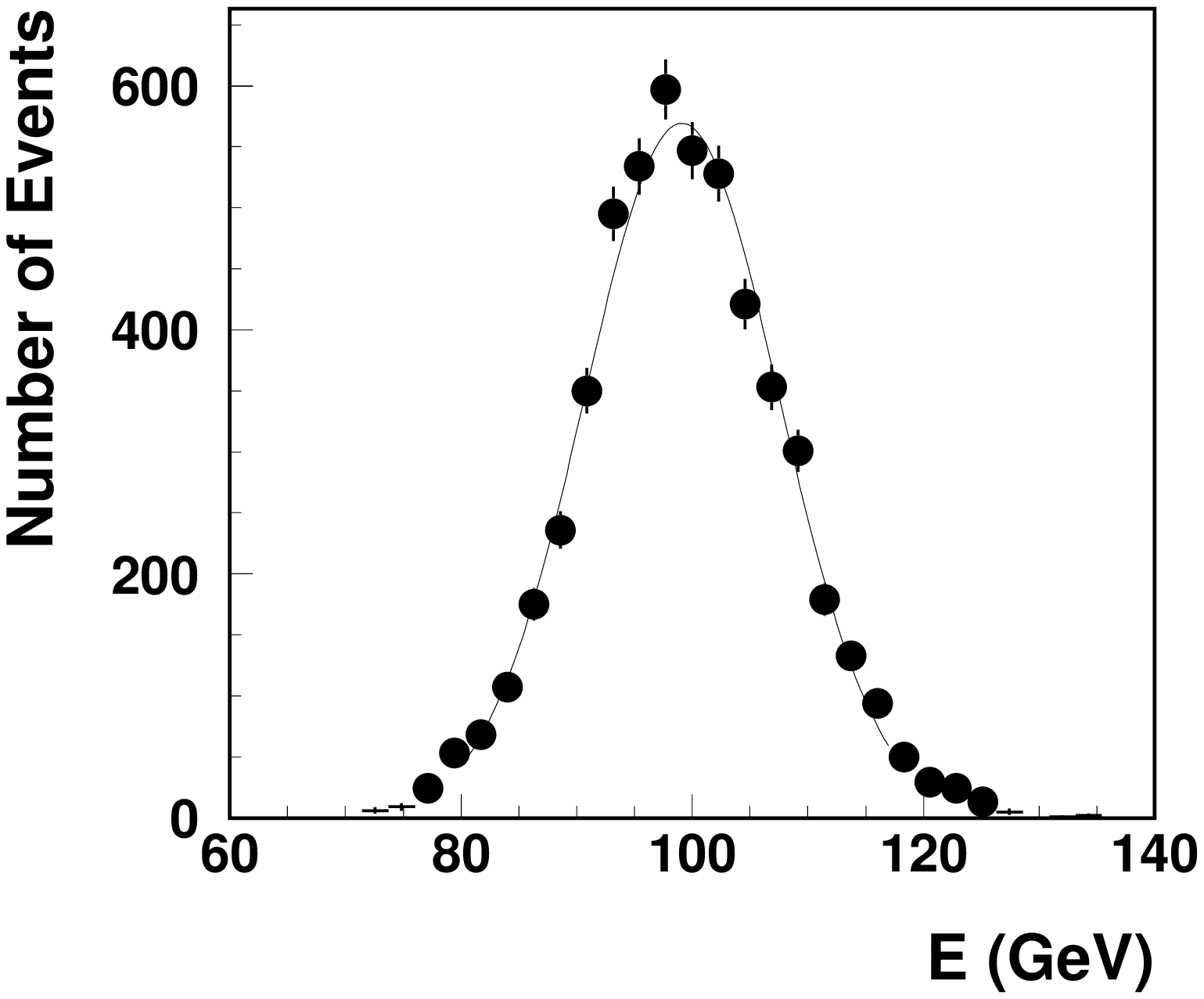,width=0.45\textwidth,height=0.4\textheight} 
\\
\epsfig{figure=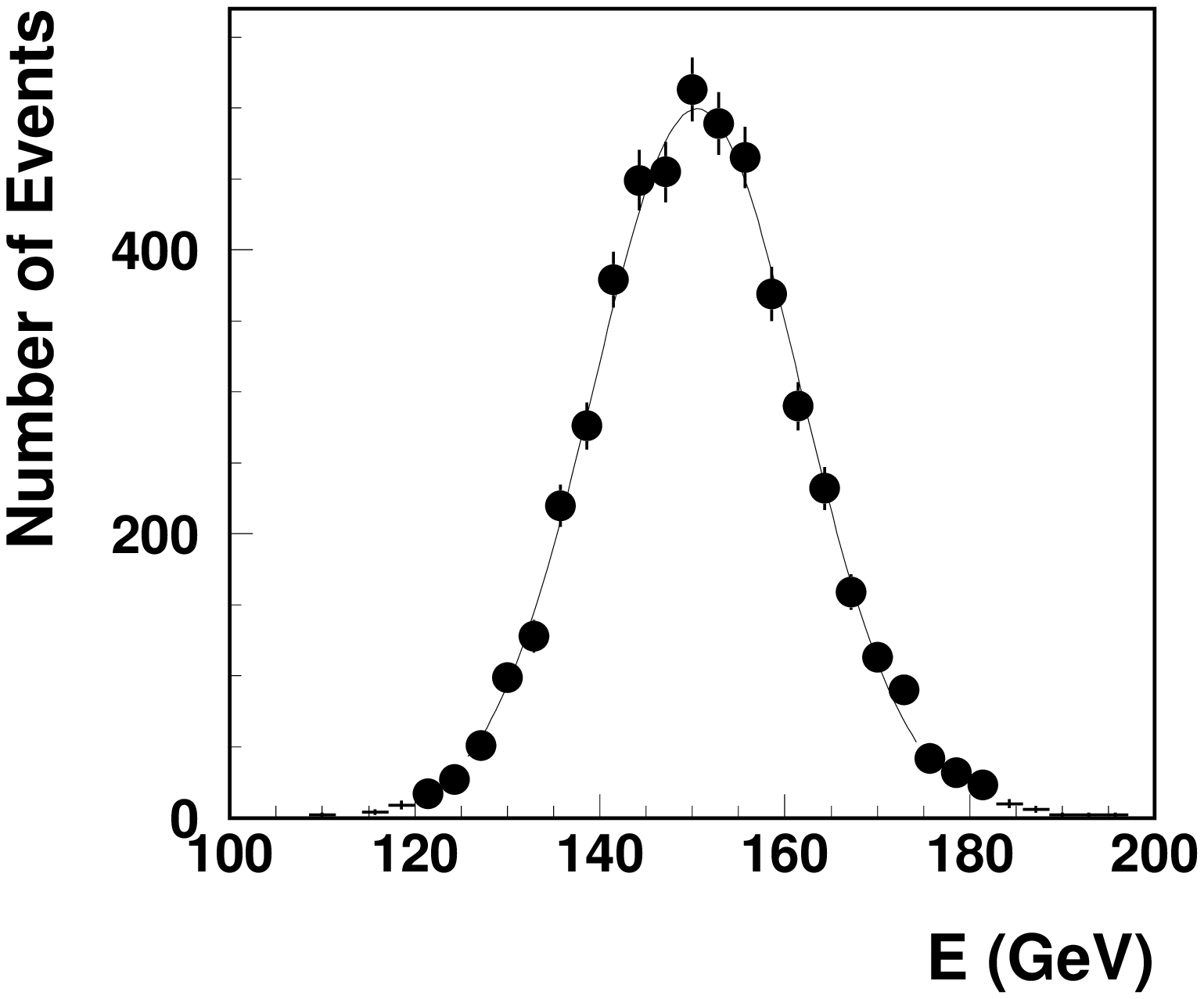,width=0.45\textwidth,height=0.4\textheight} 
&
\epsfig{figure=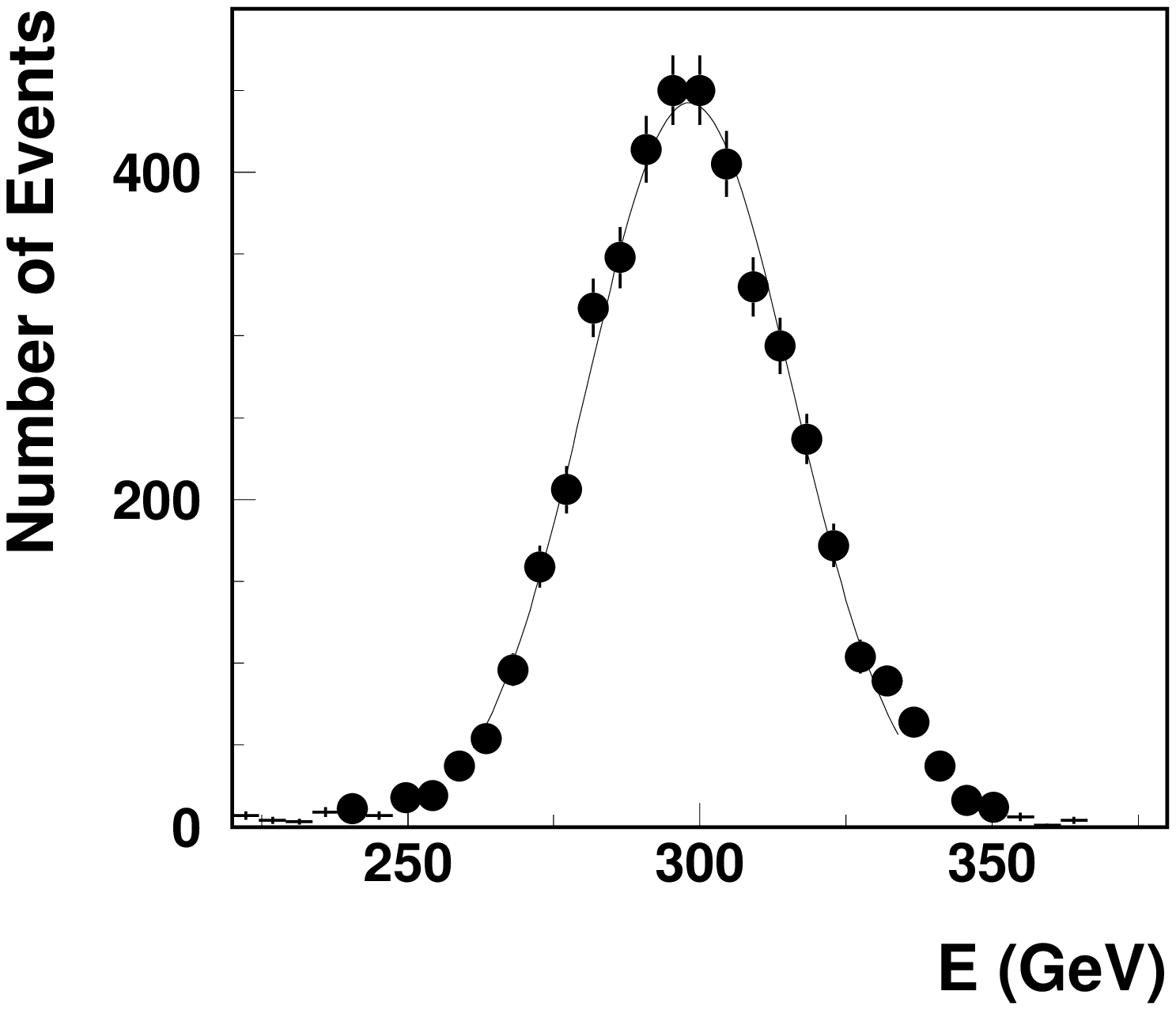,width=0.45\textwidth,height=0.4\textheight} 
\\
\end{tabular}
\end{center}
       \caption{
                The energy distributions for $E_{beam}$ = 80, 150  GeV
                (left column, up to down) 
                and $E_{beam}$ = 100, 300  GeV (right column, up to down).  
       \label{f02}}
\end{figure*}

\hspace{6mm}
Fig.\ \ref{f01} and Fig.\ \ref{f02} show the pion energy spectra reconstructed
with the $e / h$ method ($\epsilon = 0.1\%$).
The mean and $\sigma$ values of these distributions are extracted with 
Gaussian fits  over $\pm 2 \sigma$ range.
The  obtained mean values $E$, the energy resolutions $\sigma$,
and the fractional energy resolutions $\sigma / E$
are listed in Table \ref{tv1} for the various beam energies.

\subsection{Energy Linearity}
\begin{figure*}[tbph]
\begin{center}   
\begin{tabular}{c}
\epsfig{figure=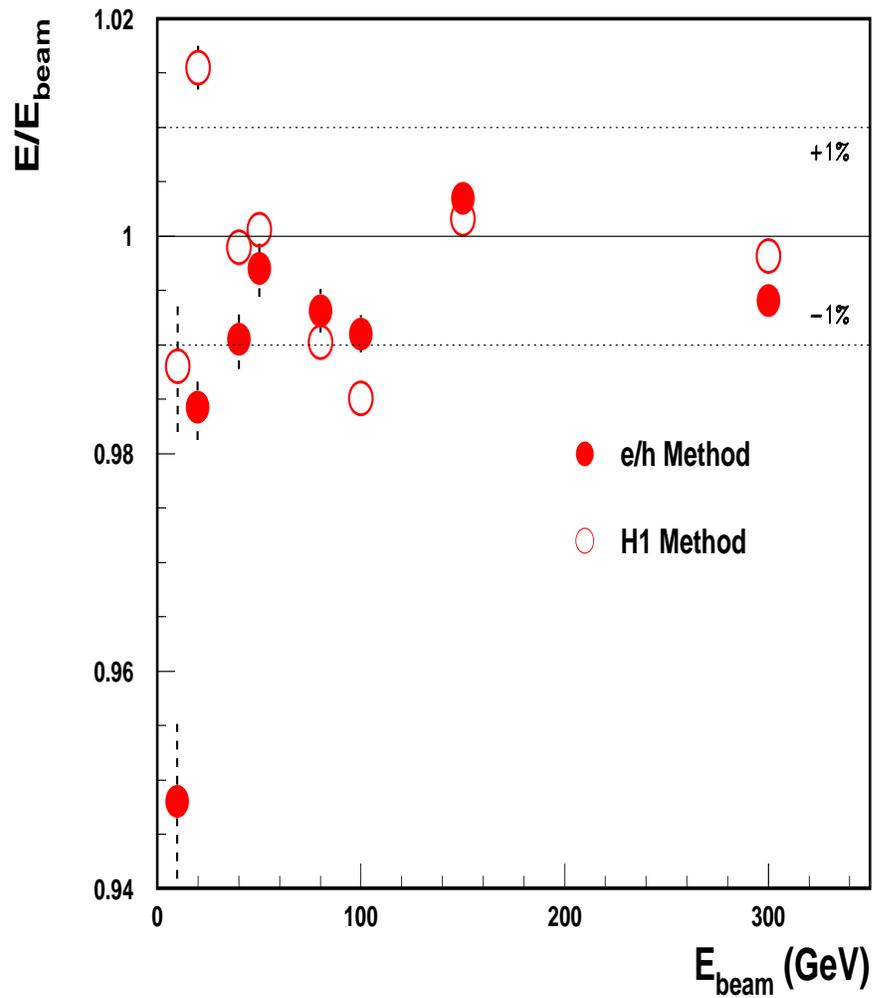,width=0.95\textwidth,height=0.9\textheight} 
\\
\end{tabular}
\end{center}
\vspace*{-20mm}
       \caption{
         Energy linearity as a function of the beam energy for 
         the $e/h$ method  
         (black circles) and the cells weighting method (open circles).
       \label{f03}}
\end{figure*}

\begin{figure*}[tbph]
\begin{center}   
\begin{tabular}{c}
\epsfig{figure=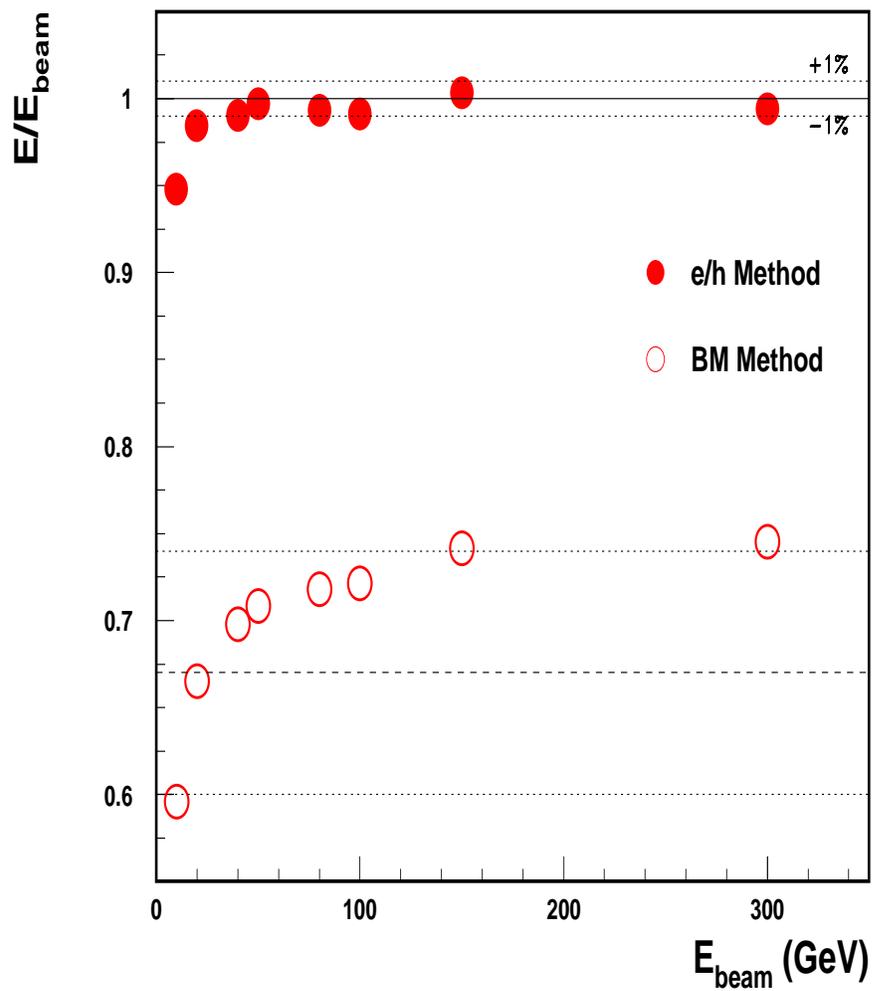,width=0.95\textwidth,height=0.9\textheight} 
\\
\end{tabular}
\end{center}
\vspace*{-20mm}
       \caption{
         Energy linearity as a function of the beam energy for 
         the $e/h$ method (black circles) and the benchmark method 
         (open circles).
       \label{f04}}
\end{figure*}

\hspace{6mm}
Fig.\ \ref{f03} demonstrates the correctness of the mean energy reconstruction.
The mean value of $E / E_{beam}$ equals to $(99.5\pm0.3) \%$ and the spread is
$\pm 1 \%$ except for the point at 10 $GeV$.
But, as noted in 
\cite{cobal98}, 
at this point the result is strongly dependent on the effective capability to 
remove events with interactions in the dead material upstream and to 
deconvolve the real pion contribution from the muon contamination. 
Fig.\ \ref{f03} and Fig.\ \ref{f04} compare the linearity, $E / E_{beam}$, 
as a function of the beam energy for the $e / h$ method,
for the cells weighting method  and for the benchmark method.
For the cells weighting method the linearity is good to about $\pm1 \%$.  
The benchmark method (without rescaling) 
\cite{comb96} 
badly reconstructs the beam energy: 
the reconstructed energy is systematically underestimated of about $30 \%$,
the linearity depends from the beam energy and the ratio $E / E_{beam}$ 
increases by $15 \%$ from 20 to 300 $GeV$. 
 
\subsection{Energy Resolutions}
\begin{figure*}[tbph]
\begin{center}   
\begin{tabular}{c}
\epsfig{figure=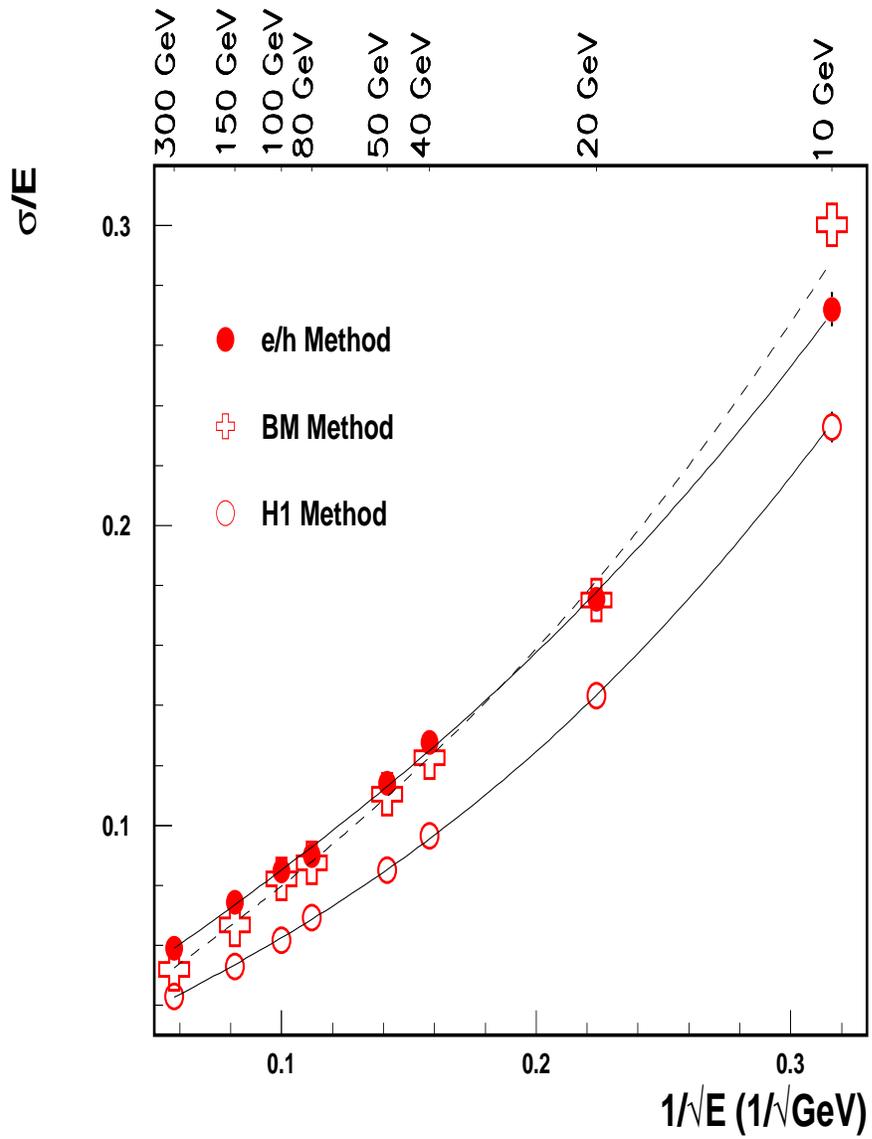,width=0.95\textwidth,height=0.9\textheight} 
\\
\end{tabular}
\end{center}
 \vspace*{-20mm}
      \caption{
         The energy resolutions obtained with the $e/h$ method (black circles),
          the benchmark method (crosses) and the cells weighting method 
         (open circles).          
       \label{f05}}
\end{figure*}

\hspace{6mm}
Fig.\ \ref{f05} shows the fractional energy resolutions ($\sigma / E$)
as a function of $1 / \sqrt{E_{beam}}$  obtained by these three methods.
As can be seen, the energy resolutions for the $e / h$ method (black circles)
are comparable with the benchmark method (crosses) and only of $30 \%$ worse 
than for the cells weighting  method (open circles).
A fit to the  data points gives the fractional energy resolution for the 
$e / h$ method obtained using the iteration procedure with $\epsilon = 0.1\%$
\begin{equation}       
        \frac{\sigma}{E} = 
                        \left[
                        \frac{(58\pm3)\%}{\sqrt{E}}
                        + (2.5\pm0.3)\% \right] 
                        \oplus \frac{(1.7\pm0.2)\ GeV}{E} \ ,
\end{equation}       
for the $e / h$ method using the first approximation (\ref{epti-00})
\begin{equation}       
        \frac{\sigma}{E} = 
                        \left[
                        \frac{(56\pm3)\%}{\sqrt{E}}
                        + (2.7\pm0.3)\% \right] 
                        \oplus \frac{(1.8\pm0.2)\ GeV}{E} \ ,
\end{equation}       
for the benchmark method of
\begin{equation}       
        \frac{\sigma}{E} = 
                        \left[
                        \frac{(60\pm3)\%}{\sqrt{E}}
                        + (1.8\pm0.2)\% \right] 
                        \oplus \frac{(2.0\pm0.1)\ GeV}{E} \ , 
\end{equation}       
and for the cells weighting  method of 
\begin{equation}       
        \frac{\sigma}{E} = 
                        \left[
                        \frac{(42\pm2)\%}{\sqrt{E}}
                        + (1.8\pm0.1)\% \right] 
                        \oplus \frac{(1.8\pm0.1)\ GeV}{E} \ ,
\end{equation}       
where the symbol $\oplus$ indicates a sum in quadrature.
As can be seen, the sampling term is consistent within errors
for the $e/h$ method and the benchmark method and is smaller
by 1.5 times for the cells weighting method.
The constant term is the same for the benchmark method and the cells 
weighting method and is larger by $(0.7\pm0.3) \%$ for the $e/h$ method.
The noise term of about $1.8\ GeV$ is the same for these three methods
that reflects its origin as the electronic noise.
As to the two approaches for the $e / h$ method the fitted parameters 
coincides within errors.

\section{Conclusions}
\hspace{6mm}
The new simple method of the energy reconstruction for a combined calorimeter,
the $e / h$ method, is suggested.
It uses only the known $e / h$ ratios and the electron calibration constants 
and does not require the previous determination of any  parameters by a 
minimization technique.
The method has been tested on the basis of the 1996 test beam data of the 
ATLAS combined prototype calorimeter and demonstrated the correctness 
of the reconstruction of the mean values of energies.
The obtained fractional energy resolution is
$[(58\pm3)\%/\sqrt{E}+(2.5\pm0.3)]\oplus (1.7\pm0.2)/E$. 
This algorithm can be used for the fast energy reconstruction in the trigger 
of the first level.

\section{Acknowledgments}
\hspace{6mm}
This work is the result of the efforts of many people from the ATLAS
Collaboration.
The authors are greatly indebted to all Collaboration for their test beam 
setup and data taking.
Authors are grateful Peter Jenni and Marzio Nessi for fruitful discussion 
and support of this work. 
We are thankful Julian Budagov and Jemal Khubua for attention and support of 
this work.





\newpage
\clearpage
\clearpage
\clearpage
\clearpage
\clearpage
\clearpage
\clearpage

\end{document}